\documentclass[12pt, twoside]{amsart}
\usepackage{inputenc}

\usepackage{amsmath}
\usepackage{amsthm}
\usepackage{amssymb}
\usepackage{amsfonts}
\usepackage[matrix,arrow,curve]{xy}

\newtheorem{theorem}{Theorem}[section]
\newtheorem{lemma}[theorem]{Lemma}
\newtheorem{proposition}[theorem]{Proposition}
\newtheorem{corollary}[theorem]{Corollary}
\theoremstyle{definition}

\newenvironment{proofof}[2]{{\itshape \noindent Proof of #1 \ref{#2}}}{\hfill\(\square\)}

\newtheorem{remark}[theorem]{Remark}
\newtheorem{note}[theorem]{Note}

\newcommand{\be}{\begin{equation}} \newcommand{\ba}{\begin{aligned}} 
\newcommand{\ee}{\end{equation}}     \newcommand{\ea}{\end{aligned}}

\newcommand{\sgn}{\text{sgn}}

\newcommand{\nc}{\newcommand}
\nc{\Symm}{{\on{Sym}}}

\newcommand{\on}{\operatorname}

\nc{\cE}{{\cal E}}

\newcommand{\fb}{{\mathfrak b}}

\nc{\SL}{{\mathfrak sl}}

\nc{\HH}{{\mathfrak h}}
\newcommand{\g}{{\mathfrak{g}}}

\newcommand{\n}{{\mathfrak{n}}}

\nc{\wh}{\widehat}\nc{\wt}{\widetilde}

\newcommand{\half}{\textstyle{\frac{1}{2}}}

\newcommand{\ben}{\begin{enumerate}}
\newcommand{\een}{\end{enumerate}}

\newcommand{\XX}{{\mathbb{X}}}

\newcommand{\RR}{{\mathbb R}}
\newcommand{\NN}{{\mathbb N}}

\newcommand{\PP}{{\mathbb{P}}}

\newcommand{\ZZ}{{\mathbb{Z}}}

\newcommand{\cC}{{\mathcal C}}

\newcommand{\cD}{{\mathcal D}}

\newcommand{\cW}{{\mathcal W}}

\newcommand{\cinf}{C^\infty}

\newcommand{\cinfty}{C^\infty}

 \usepackage{hyperref}

\begin{document}

\title[Poisson reduction of the space of polygons]
{Poisson reduction of the space of polygons}

\author{Ian Marshall}
\address{Mathematics Department, University of Loughborough, UK}
\email{ian.marshall@unige.ch}

\maketitle

\begin{abstract}
A family of Poisson structures, parametrised by an arbitrary odd periodic function $\phi$, is defined on the space $\cW$ of twisted polygons in $\RR^\nu$. Poisson reductions with respect to two Poisson group actions on $\cW$ are described. 
The $\nu=2$ and $\nu=3$ cases are discussed in detail and the general $\nu$ case in less detail.
Amongst the Poisson structures arising in examples are to be found the lattice Virasoro structure, the second Toda lattice structure and some extended Toda lattice structures. A general result is proved showing that, for any $\nu$, to certain concrete choices of $\phi$ there correspond compatible Poisson structures which generate all the extended bigraded Toda hierarchies of a suitable size.
\end{abstract}

\section*{Introduction}

The space ${\mathcal D}_\nu$ of scalar differential operators $L$ of order $\nu$ with 
periodic coefficients,
\be\label{defD}
{\mathcal D}_\nu\owns L= \partial^\nu + u_{\nu-1}\partial^{\nu-1} + \cdots + u_1\partial + u_0,\qquad \partial = d/dx,\ \ u_k\in\cinfty(\RR/2\pi\ZZ,\RR),
\ee
is a Poisson subspace, with respect to two compatible Poisson structures, of the space of periodic pseudo-differential operators. These Poisson structures are usually associated with the names of Adler and of Gelfand-Dikii, see \cite{adler,gd}. Alternatively ${\mathcal D}_\nu$ may be subjected to analysis via a Hamiltonian reduction procedure \cite{ds}, known in this context as \emph{Drinfeld-Sokolov reduction}, which results in the same pair of Poisson structures. The reduction procedure rests on the observation that in writing the 
equation $L\psi=\psi$ 
as a $\nu\times\nu$ matrix system $\partial\Psi + {\mathcal L}\Psi=0$, the matrix $\mathcal L$ is not uniquely defined. Let 
$\fb=\{$lower triangular matrices$\}$ and $\n:=\{$strictly lower triangular matrices$\}$. 
$\fb$ and $\n$ are Lie subalgebras of $gl_\nu$ with associated Lie groups 
$B=\{$lower triangular matrices$\}\cap GL_\nu$ and $N=\{A+Id|A\in\n\}$. 
Defining the matrix $\Lambda = \sum_{i=1}^{\nu-1}E_{i,i+1}$
$$
\Lambda = \begin{pmatrix}			
      0        &1           &\cdots&0        \\		
      \vdots&\ddots& \ddots &\vdots\\		
      \vdots& \cdots& \ddots &1\\			
       0     & \cdots& \cdots &0			
\end{pmatrix},
$$
then $A\in\fb$ may be chosen so that 
${\mathcal L}=-\Lambda - A$. A canonical choice for $A$ is 
$$
A_{can}=
\begin{pmatrix}
      0        &\cdots& 0        \\
      \vdots&           &\vdots\\
      0        &\cdots & 0        \\
      u_0   & \cdots   &u_{\nu-1}
\end{pmatrix},
$$
corresponding to $\Psi_{can}^T = (\psi,\psi',\cdots,\psi^{(\nu-1)})$. An equivalent representation of
$L\psi=0$ 
can be obtained by making a different choice for $\Psi$. Thus  $\Psi$ may be
replaced by $\hat\Psi=g\Psi_{can}$ for some $g\in N$ whose entries are periodic functions, to get 
$\partial\hat\Psi + \hat{\mathcal L}\hat\Psi=0$. In other words, we have 
 a gauge freedom in the passage from $L$ to $\mathcal L$ via the action of the nilpotent subgroup 
 $N$ of strictly lower triangular matrices: ${\mathcal L}\sim\hat{\mathcal L} \Leftrightarrow\exists g\in N$ s.t. $\hat{\mathcal L} = g{\mathcal L}g^{-1}+(\partial g)g^{-1}$. 
 That is ${\mathcal D}_\nu = \{{\mathcal L}=-\Lambda + A\ |\ A\in\fb\}/N$. 
 The key to the Drinfeld-Sokolov reduction procedure is the identification of a Poisson structure 
 on the space of $\mathcal L$s with respect to which the gauge action is Poisson. In fact this is the 
 standard \emph{Lie-Poisson} structure on the dual of the central extension of the loop algebra of 
 $gl_\nu$. Fixing the form of $\mathcal L$ to be ${\mathcal L}=-\Lambda + A$ with $A$ a 
 lower-triangular matrix is the fixing of the momentum map for the gauge action.
The construction may be extended to an arbitrary semi-simple 
 Lie group $G$, replacing the role of $\Lambda$ by a principal nilpotent element in $\g=Lie(G)$, 
 and by replacing the group of lower triangular matrices by the appropriate Borel subgroup and the 
 strictly lower ones by the appropriate nilpotent subgroup.

An analogous construction was undertaken in the articles \cite{frs} and \cite{ss} for $q$-difference 
operators and for shift operators on a one-dimensional lattice. From now on everything will be in the setting of the one-dimensional lattice, although most, if not all, of what follows here 
applies equally to the $q$-difference case.\footnote{It would be of some interest - at least to this author - to find the $q$-difference version of (\ref{PBonW}).}
One would like to discover a Poisson structure on the space 
${\mathcal S}_\nu$ of scalar shift-operators $L$ of order $\nu$ with periodic coefficients,
\be\label{defS}
\ba
{\mathcal S}_\nu\owns L= D^\nu + u_{\nu-1}D^{\nu-1} + \cdots + u_1D + u_0,&\\ 
(Df)_m=f_{m+1}\ \hbox{for}\ f\in Fun(\ZZ,\RR)&=\ \hbox{infinite sequences in $\RR$},\\
u_k\in Fun(\ZZ/N\ZZ,\RR)&=\ \hbox{periodic sequences in $\RR$}. 
\ea
\ee
Exactly the same argument as described above for the operators $L\in\cD_\nu$ in (\ref{defD})
holds for the operators $L\in{\mathcal S}_\nu$ in (\ref{defS}), except that now the freedom in the 
choice of representation of the scalar equation is generated by $\hat\Psi=g\Psi$ for $g\in N$ 
with entries in $Fun(\ZZ/N\ZZ,\RR)$. One arrives at the point where to cast the reduction procedure 
in a Poisson setting, a Poisson structure is required on the space of $\mathcal L$s, with 
respect to which the gauge-action $\hat{\mathcal L}=(Dg){\mathcal L}g^{-1}$ be Poisson: this is 
provided by the one due to Semenov-Tian-Shansky, given in \cite{stsrims} the label \emph{twisted 
lattice current algebra}. Hence the natural setting for the Poisson description of difference 
operators is that of Poisson Lie groups and it involves the choice of a suitable $r$-matrix. The 
construction may also be extended to the cases for which $GL_\nu$ is replaced by arbitrary 
semisimple Lie groups.

The Poisson algebras resulting from the Drinfeld-Sokolov setting are known, in the language of 
Conformal Field Theory, as W-algebras. Those resulting from the 
Frenkel-Reshetikhin-SemenovTianShansky-Sevostyanov setting are known by analogy as 
$q$-deformed or lattice W-algebras. The simplest example of a W-algebra is the Virasoro algebra, which comes from the $sl_2$ case of reduction. The corresponding lattice W-algebra is called, by analogy, the \emph{lattice Virasoro algebra}. An equivalent lattice analogue of Virasoro is the \emph{Faddeev-Takhtajan-Volkov} algebra. As was described in \cite{frs} there is a map between the two algebras. This is discussed in Section 2 of the present article.

In both the Drinfeld-Sokolov setting for differential operators and the 
Frenkel-Reshetikhin-SemenovTianShansky-Sevostyanov setting it is perceived  -- either as 
a useful observation in the DS setting, or as a crucial requirement in the FRS-SS setting -- that 
fixing the form of $\mathcal L$ to ${\mathcal L}=\Lambda+A$, with $A$ 
lower-triangular, is a constraint of \emph{first class type}. Indeed this was invoked in \cite{frs} and 
in \cite{ss} as a device by means of which the exact form of the $r$-matrix they needed was to 
be fixed uniquely.

In the present article there will be presented an alternative way to obtain a Poisson structure on 
the space ${\mathcal S}_\nu$ of shift operators. The computations will be presented in fullest 
detail only  for second and third order examples, but this is enough to illustrate the principal
idea which, as in the articles \cite{frs, ss}, is to first identify the space 
${\mathcal S}_\nu$ as a quotient space of a more elementary space, and then to interpret the
projection from the elementary space to the quotient as a reduction using a standard 
argument from the theory of Poisson groups. The most striking  thing is the indication that the 
requirement in \cite{frs} and \cite{ss} for constraints to be first-class was unnecessarily strong, and 
that by relaxing it the results which may be incorporated in the reduction procedure are extended.

To justify this last claim, consider the space of second order difference operators on a 
periodic lattice of length $N$,
\be\label{order2L}
L= D^2 -uD + \rho,\qquad u,\rho\in Fun(\ZZ/N\ZZ,\RR),\ \  (Df)_n=f_{n+1}.
\ee
On the one hand the space of such operators is familiar as being the same as the set of periodic tri-diagonal matrices, providing the standard setting for the Toda lattice system. On the other hand, after fixing $\rho\equiv1$ it is a discrete analogue of the space of periodic Schr\"odinger operators, on which the standard Poisson structure - as found in \cite{frs} - is identified with the Faddeev-Takhtajan-Volkov \cite{ft,v} Poisson structure; itself a discrete analogue of the Virasoro algebra, which is in turn naturally related to standard periodic Schr\"odinger operators, $L=\partial^2 +u$. 

The point of view described in the present article admits an element of freedom absent in \cite{frs},
allowing the recovery not only of the lattice Virasoro structure upon constraining $\rho\equiv1$ in 
(\ref{order2L}), but also of the ``second Toda lattice Poisson bracket''  without fixing $\rho$. 
One begins with the space 
$\cW$ of quasi-periodic sequences in $\RR^2$ on which a family of Poisson structures is defined, 
parametrised by an \emph{arbitrary} odd function $\phi$ and on which the action of 
$GL_2$ is a Poisson action. The space of operators of the form given in (\ref{order2L}) is the 
same as $\cW/GL_2$. The function $\phi$ may be viewed as a remnant of the initial freedom (subsequently 
relinquished) in the definition of the $r$-matrix in \cite{frs,ss}. For one choice for $\phi$, which 
corresponds to the fixing of the $r$-matrix imposed in \cite{frs,ss}, $\rho\equiv1$ is a first-class 
constraint 
and one obtains the lattice Virasoro algebra directly. For another choice, which 
produces the Toda Poisson structure, $\rho\equiv1$ is not a constraint of first-class type, but 
reduction by the Dirac method can be made nonetheless. A simple corollary, which follows from 
Proposition \ref{projprop}, is that constraining of the second Toda Poisson bracket - by the 
standard Dirac method - yields the lattice Virasoro structure. This fact appears not to have been 
previously published, although it was not unknown to experts \cite{d}. Here it is given an 
explanation of sorts.

It is appropriate to make the remark here that the constraint on $\rho$ may be more general;
that is, for any fixed periodic $\beta$, by imposing $\rho\equiv\beta$. The result is a 
generalisation of the lattice Virasoro algebra in which $\beta$ furnishes a set of parameters. This 
more general Poisson structure appeared -- apparently for the first time -- in the article \cite{vs}.

The results presented in this article are closely related to those of the article \cite{ms}. 
The main idea is to start with the exchange algebra defined in (\ref{PBonW}) and in Proposition \ref{basic}. This was also the point of departure in the article \cite{Bab} of Babelon which treated similar questions to the ones dealt with here.

\bigskip

At various stages this work was done during a number of visits to different universities and research institutes, whilst I was wandering around with no job. I wish to express my gratitude to all of my hosts, sometimes for their hospitality and encouragement, sometimes for their help and suggestions and in many cases for all of these.

\section{Poisson structure on the space of polygons in $\RR^\nu$}

Let $N\in\NN$ be fixed. We consider the space of twisted polygons of length $N$ in $\RR^\nu$, 
which will be denoted $\cW$. (It is assumed that $N$ is reasonably large compared with $\nu$. If 
not, then some of the arguments which follow later do not make sense.)

\begin{note}\emph{In this article, elements in $\RR^\nu$ are always written as row-vectors, 
except within determinants, where they will appear as column vectors.}
\end{note}

An element of $\cW$ is a pair $(V,M)$, where $V:\ZZ\rightarrow\RR^\nu\backslash\{\mathbf0\}$ is a sequence in 
$\RR^\nu\backslash\{\mathbf0\}$ and $M$ is an element in $GL_\nu$. $V$ and $M$ are related by the condition of 
\emph{quasi-periodicity} or \emph{twisting}, $V_{n+N}=V_nM\quad\forall n$. 
Thus we define
\be\label{DefW}
\cW=\{(V,M)\in Fun(\ZZ,\RR^\nu\backslash\{\mathbf0\})\times GL(\nu,\RR)|\ V_{k+N}=V_kM\ \forall n\}.
\ee
There are two natural group actions on $\cW$. Introduce the group 
$\cC=Fun(\ZZ/N\ZZ,\RR^\times)$ of periodic sequences of non-zero real numbers, with 
$(pq)_m=p_mq_m$ for $p,q\in\cC$. The groups $GL_\nu$ and $\cC$ both have natural (commuting) actions 
on $\cW$, 
\be\label{GroupActions}
\ba
(p,g)\cdot(V,M) = (pVg^{-1}, gMg^{-1})\qquad p\in\cC, g\in GL_\nu.
\ea
\ee
The two group actions are not disjoint. For $k\in\RR^\times$, the actions of $k\in\cC$ and 
$k^{-1}\,Id\in GL_\nu$ have the same effect. To resolve this it is convenient to replace $GL_\nu$ by $SL_\nu$ and to 
leave $\cC$ intact. It is also natural to restrict $M$ to lie in $SL_\nu$. Denote $SL(\nu,\RR)$ by 
$G$ and $sl(\nu,\RR)$ by $\g$. 

Let $R\in\wedge^2\g$ and let $C\in S^2\g$. Set $R_\pm=\half(R\pm C)$. Let 
$\phi\in Fun(\ZZ/N\ZZ,\RR)$ be an odd periodic function. Denote by $\sigma$ the ``discrete 
sign function'', $\sigma_m= \sgn(m)$ if $m\neq0$ and $\sigma_0=0$.
Define the bracket on $\cW$
\be\label{PBonW}
\left\{\begin{array}{llll}
&\{V_m^1,V_n^2\}&= &V_m\otimes V_n[R+\sigma_{m-n}(C+Id\otimes Id) + \phi_{m-n}],\\
&\{V_m^1,M^2\}&= &V_m^1[M^2R_- - R_+M^2],\\
&\{M^1,M^2\}&= &(M\!\otimes\!M)R + R(M\!\otimes\!M) - M^1R_+M^2 - M^2R_-M^1.
\end{array}\right.
\ee
\begin{proposition}\label{basic}
The formulae in (\ref{PBonW}) define a Poisson bracket on $\cW$ if and only if $C$ is the Casimir element and if $R$ is a classical r-matrix, i.e. $R$ satisfies the Yang-Baxter equation 
$[R^{12},R^{13}] + c.p. = - [C^{12},C^{13}]$.
\end{proposition}

\begin{proof}
The proof is a straightforward check of the Jacobi identity.
\end{proof}
Note that $C$ is defined by the property $(g\otimes h)C= C(h\otimes g)\ \forall\, g,h\in G$. Alternatively, 
\be\label{Cexprop}
(\xi\otimes\eta)(C+Id\otimes Id) = \eta\otimes\xi\qquad \forall\xi,\eta\in\RR^\nu.
\ee
$R$ defines the structure of a Poisson Lie group on $G$ by the Sklyanin formula
$$
\{g^1,g^2\}= Rg\otimes g - g\otimes gR.
$$

\begin{proposition}
The actions of $\cC$ and $G$ given by (\ref{GroupActions}) are Poisson actions in the Poisson Lie 
group sense, when $G$ has the Sklyanin Poisson structure and $\cC$ has the zero Poisson 
structure.
\end{proposition}

Primes denote shifts of order up to 3 and a superscript with the corresponding 
number in brackets denotes higher order shifts, thus for $f\in Fun(\ZZ,\RR)$, 
\be\label{defprime}
\ba
f':=(Df)\quad&\hbox{and}\quad 
f^{(r)}:=(D^rf),\\
\hbox{or}\qquad (f')_m:=f_{m+1}\quad&\hbox{and}\quad 
(f^{(r)})_m:=f_{m+r}.
\ea
\ee
Define the Wronskian $w:\cW\rightarrow\RR$ by
\be\label{defWronskian}
w(V,M)=|VV'\dots V^{(\nu-1)}|, \quad\hbox{i.e.}\ w_m=|V_mV_{m+1}\dots V_{m+\nu-1}|.
\ee
Usually it will be convenient to write $w(V)$ instead of $w(V,M)$.

\begin{proposition}\label{momprop}
The map $(V,M)\mapsto M$ is a momentum map for the action of $G$ (in the Poisson Lie group 
sense, see \cite{l}). The map $(V,M)\mapsto w(V,M)$ is a momentum map for the action of $\cC$.
\end{proposition}

\begin{proof}
The first claim follows directly from the second and third formulae in (\ref{PBonW}). The second is an interpretation of the formula
\be
\{w_m,V_n\}=\left(\sigma_{n-m}+\sum_{r=0}^{\nu-1}\phi_{m+r-n}
+\sum_{r=1}^{\nu-1}\delta_{m+r-n}\right)w_mV_n,
\ee
which is proved in the appendix.
\end{proof}

\begin{remark}
The space $\cW$ was introduced in \cite{ms}, where it was observed that there were natural 
actions of the groups $\cC$ and $G$ on $\cW$. 
As was explained in that article, the Poisson structure may be determined completely, up to the 
freedom in the choice of the arbitrary odd periodic function $\phi$, just by the requirement that the 
actions of $\cC$ and $G$ on $\cW$ be Poisson. 
The choice of label $\cW$ is inherited from the role this space plays as the space of ``wave-functions'' when it arises as solutions of some family of linear problems. It is an important and subtle distinction that here $\cW$ is viewed as an independent entity and that the linear problems should rather be seen as being generated from $\cW$ than the other way about. 
\end{remark}
It 
follows from Poisson Lie theory that the set $C^\infty(\cW)^\cC$ of smooth 
$\cC$--invariant functions on $\cW$ is closed with respect to the Poisson bracket 
(\ref{PBonW}), hence that a Poisson structure is automatically induced on the quotient space 
$\cW/\cC$ and the projection $\cW\rightarrow\cW/\cC$ is a Poisson map.

Projection from $\cW$ to $\cW/\cC$ is essentially the projection from 
$\RR^\nu\backslash\{{\mathbf0}\}$ to 
$\RR \PP^{\nu-1}$. On the open subset of $\RR^\nu$ in which the last component is non-zero, 
we may define the projection from $\cW$ to $\cW/\cC$ in coordinates by first writing 
$\RR^\nu\owns V =\chi(v,1)$ with $v\in\RR^{\nu-1}$ and $\chi\in\RR^\times$. Then the map is given by 
$\chi(v,1)\mapsto v$. 

In \cite{ms} the main focus of interest was in applying Poisson Lie group reduction to obtain 
a sequence of Poisson projections starting with $\cW$ and ending (for the case $\nu=2$) with the 
lattice Virasoro algebra. Here the emphasis will be slightly different and the following result, valid for arbitrary $\nu$, will be an important one. 

\begin{proposition}\label{projprop}
The Poisson structure obtained by projection $\cW\rightarrow\cW/\cC$ has the form
\be\label{projPB}
\{v_m^1,v_n^2\}=(v_m\otimes v_n)\cdot R -
\sigma_{m-n}(v_m-v_n)\otimes(v_m-v_n).
\ee
\end{proposition}

\begin{proof} See appendix.
\end{proof}

Here the notation $(a\otimes b)\cdot R$ denotes the 
projective action of $R$ on $(\RR\PP^{\nu-1}\otimes\RR\PP^{\nu-1})$.
Formula (\ref{projPB}) must be complemented by suitable representations of $\{v^1_m,M^2\}$ and $\{M^1,M^2\}$, but these will not be needed, so they are not written here. Two properties of the Poisson structure on $\cW/\cC$ are:

\begin{enumerate}
  \item The action of $G$ on $\cW/\cC$ defined by projectivising the action of $G$ on $\cW$ is a Poisson action.
  \item A striking and noteworthy feature of formula (\ref{projPB}) is that it does not depend on $\phi$. Crucial use will be made of this in the next sections.
\end{enumerate}
\begin{remark}
From now on we restrict ourselves to the open set of \emph{non-degenerate} elements in $\cW$, 
for which $w(V,M)\neq0$.
\end{remark}

The following result\footnote{
I am grateful to G.Falqui for teaching me this result, which clearly ``everybody should know''.}
justifies the subsequent interest in ``\emph{Poisson tensors with a linear 
dependence on one of the variables}''

\begin{proposition}
\label{GF}
Let $M$ be a Poisson manifold with Poisson tensor $P$. Let $\xi$ be a vector field on $M$. 
Suppose that $(L_\xi)^2P=0$. Then $L_\xi P$ is also a Poisson tensor, from which it follows that 
the pair $(P,L_\xi P)$ defines a bi-hamiltonian structure on $M$.
\end{proposition}
\begin{proof}
As $P$ is Poisson, we have, using the Schouten bracket notation, \([P,P]=0\). Applying $L_\xi$ to this condition,
\be\label{compatibility}
0=L_\xi[P,P] = 2[L_\xi P,P].
\ee
Applying $L_\xi$ again,
\be\label{poisson}
0=L_\xi[L_\xi P,P]=[(L_\xi)^2P,P] + [L_\xi P, L_\xi P] = [L_\xi P, L_\xi P].
\ee
It follows from (\ref{poisson}) that $L_\xi P$ is Poisson and it follows from (\ref{compatibility}) that
$P$ and $L_\xi P$ are compatible.
\end{proof}

It is worth also giving a qualification of the notion of linearity of a Poisson tensor. Later in this 
article we look at Poisson structures which are quadratic-plus-linear in terms of the 
variables chosen to present them. This means some local coordinates $\mathbf x$ are chosen 
for which all Poisson brackets have the form
\[
\{x^i,x^j\}={\mathbf x}^TA(i,j){\mathbf x} +{\mathbf x}^T{\mathbf b}(i,j)= \sum_k\sum_lA(i,j)_{kl}x^kx^l
+\sum_k b(i,j)_kx_k,
\]
where for each $(i,j)$, $A(i,j)$ is a constant, symmetric matrix and ${\mathbf b}(i,j)$ is a constant vector. When such a Poisson tensor is said to have a linear dependence on $x^k$ for some $k$, it is meant that the $k$th 
diagonal entry of each matrix $A(i,j)$ is zero, i.e. that for all $(i,j)$, $A(i,j)_{kk}=0$, and then it is 
easy to see that the vector field $\xi$, defined by $dx^l(\xi)=\delta^l_k$, satisfies the condition of 
Proposition \ref{GF}.

\subsection{Operator notation for sequences in $\RR$}

It proves useful later on to make use of a notation based on the notion that a sequence $K\in Fun(\ZZ,\RR)$ can be viewed as the kernel of an operator on $Fun(\ZZ,\RR)$. Let $K_m\in\RR$ be a sequence of real numbers, then the corresponding operator, also denoted $K$, is defined by the following property
\be\label{opdef}
\forall f\in Fun(\ZZ,\RR),\quad (K\cdot f)_m=\sum_nK_{m-n}f_n.
\ee
Using this notation, for example, the kernel of the shift operator $D$ is the function $m\mapsto D_m=\delta_{m+1}$. The operator notation will be used extensively to treat various conditions on the function $m\mapsto\phi_m$ involved in the definition of the Poisson bracket (\ref{PBonW}).

\section{Polygons in $\RR^2$}\label{nu2}

For $V$ a  sequence in $\RR^2$, clearly, for any $m$, the three vectors 
$V_m, V_{m+1}, V_{m+2}$ are linearly dependent. Assuming that $V$ is a non-degenerate sequence, that is to say $\forall n, w_n\neq0$, the linear 
dependence may be written in the form $V_{m+2}=\mu_mV_{m+1}-\rho_mV_m$. In the more compact notation defined in (\ref{defprime}),
\be\label{lindep2}
V''=\mu V' - \rho V.
\ee
Moreover, the sequences $\mu$ and $\rho$ evidently depend on the sequence $V$ and this dependence may be given explicitly:
\be\label{murho}
w(V)\mu(V)=|VV''|,\quad w(V)\rho(V)=w(V)'.
\ee
It is straightforward to check that $\mu$ and $\rho$ are periodic, i.e. 
$\forall m\ (\mu_{m+N},\rho_{m+N})=(\mu_m,\rho_m)$, and that 
$\forall g\in G\ (\mu(Vg),\rho(Vg))= (\mu(V),\rho(V))$. Hence, as dim$(\cW)=2N+3$ and dim$(G)=3$, $\mu$ and $\rho$ are good coordinates on $\cW/G$.

An alternative set of coordinates may be defined as follows. For $V$ a quasi-periodic sequence, 
let $\Gamma(V)\in\cC$ be such that $\tilde V=\Gamma V$ has constant Wronskian, 
i.e. $w(\tilde V)\equiv const.$ Then $\tilde V'' = u\tilde V' - \tilde V$, where the coefficient of 
$-\tilde V$ is necessarily 1. If $N$ is odd $\Gamma$ is determined uniquely, whilst for $N$ even, 
although $\Gamma$ is not defined uniquely, it still exists. Meanwhile $V\mapsto u(V)$ is an invariant of the 
$\cC$-action as well as of the $G$ action and so projection from $\cW/G$ to $\cC\backslash\cW/G$ 
is implemented directly in the coordinates $(u,\Gamma)$ by forgetting $\Gamma$. The simplest 
way to implement this is to make the change of variables 
$(u,\Gamma)\mapsto (S,\gamma)=(uu',\Gamma\Gamma')$. This results in the expressions
$$
\gamma(V)=\frac{1}{w(V)}\quad\hbox{and}\quad S(V)=\gamma\gamma''|VV''|\,|VV''|'
=
\frac{|VV''|\,|VV''|'}{|VV'|\,|VV'|''}
$$
for $\gamma$ and $S$, by means of which the Poisson structure on $\cW/G$ may be explicitly computed in terms of $(S,\gamma)$. As $w$ generates the action of $\cC$ on $\cW$,
$\{\gamma_m,S_n\}$ is zero for all $m$ and $n$, so the Poisson bracket is diagonal in the $(S,\gamma)$ representation. Hence the Poisson projection from $\cW/G$ to 
$\cC\backslash\cW/G$ is represented by the algebra $\{S_m,S_n\}$ and by simply ``forgetting'' the contribution of $\gamma$.

Whilst this route could be taken and a lattice analogue of the Virasoro algebra could be obtained 
directly at this stage, this will not be the strategy pursued here, as it would be to miss another 
interesting aspect of the Poisson analysis of the reductions of $\cW$. Some intermediate 
comments are appropriate however:
\begin{enumerate}
\item It has already been pointed out that even for the case of general $\nu$, the Poisson structure 
obtained by the projection $\cW\rightarrow\cW/\cC$ is independent of the arbitrary function $\phi$. 
It follows that the Poisson structure on $\cC\backslash\cW/G$ must be independent of $\phi$.\\
\item Although it was convenient in the previous paragraph to propose the presentation of the Poisson structure on $\cC\backslash\cW/G$ via the variables $(S,\gamma)$, it is evident that the same splitting phenomenon will take place in the variables $(u,\Gamma)$ and so a representation of the reduced Poisson algebra will also exist in terms of the variables $u$. Indeed if $N$ is odd the map $u\mapsto S$ is invertible and such a representation may be obtained directly from the formula for $\{S_m,S_n\}$.
\end{enumerate}

\subsection{Reduction via the Dirac method}

Having observed that the reduction $\cW\rightarrow\cC\backslash\cW/G$ is a Poisson 
reduction and that it can be presented explicitly by computing the Poisson brackets in terms of the
coordinates $(S,\gamma)$ on $\cW/G$, it is interesting to perform the reduction by a different 
route, using the Dirac constraint method with the variables $(\mu,\rho)$ on $\cW/G$. Of course 
$\mu\vert_{\rho\equiv1}$ is the same as $u$.

Working in terms of the variables $(\mu,\rho)$ the relation in (\ref{lindep2}) is recognised to be
the standard Lax-matrix for the Toda system. More specifically, it will be the periodic Toda lattice 
due to the periodicity inherent in the setting chosen here. We undoubtedly are about to uncover 
Poisson structures associated with the same space and might expect that they have something 
to do with the Toda lattice. As we shall see, this turns out indeed to be the case.

Using the explicit expressions in (\ref{murho}), the representation of the Poisson
structure on $\cW/G$ is computed explicitly in terms of $(\mu,\rho)$. The result is the following algebra
\be\label{murhoPB}
\left\{\ba
\{\mu_m,\mu_n\} &= \Bigl(2\phi_{m-n} - \phi_{m-n+1}-\phi_{m-n-1}\\ 
&\qquad\qquad
- \delta_{m-n+1} + \delta_{m-n-1}\Bigr)
\mu_m\mu_n
+
2\delta_{m-n+1}\rho_n - 2\delta_{m-n-1}\rho_m,\\
\{\mu_m,\rho_n\} &= \Bigl(\phi_{m-n}+\phi_{m-n+1}  - \phi_{m-n-1} - \phi_{m-n+2}\\
&\qquad\qquad
- \delta_{m-n}
+ \delta_{m-n-1} +\delta_{m-n+1} - \delta_{m-n+2}\Bigr)\mu_m\rho_n,\\
\{\rho_m,\rho_n\} &= \Bigl(2\phi_{m-n}-\phi_{m-n+2} - \phi_{m-n-2}\\ 
&\qquad\qquad
- \delta_{m-n+2}+\delta_{m-n-2}\Bigr)\rho_m\rho_n.
\ea\right.
\ee
This Poisson structure depends on $\phi$, but we know in advance that the structure obtained by constraint to $\rho\equiv1$ will not depend on $\phi$. In fact there are two
special choices for $\phi$ which present themselves as deserving attention. One choice is useful for extracting the formula for the Poisson structure on $\cC\backslash\cW/G$ in terms of $u=\mu|_{\rho\equiv1}$, rather than in terms of $S$, reproducing the result of \cite{frs}. The other choice will give rise to the Poisson structure of the Toda lattice.

\subsubsection{$\phi$ chosen to diagonalise the Poisson structure} In applying the Dirac constraint method it will be convenient if we are able to make use of the freedom in the choice of $\phi$ in order to make the bracket $\{\mu_m,\rho_n\}$ identically zero. The constrained Poisson bracket will then be given by the formula for $\{\mu_m,\mu_n\}$ in (\ref{murhoPB}) with $\phi$ replaced by this special choice. The essential question here is \emph{Does there exist a solution to the following difference equation?}
\be\label{funeq1}
\phi_m+\phi_{m+1}  - \phi_{m-1} - \phi_{m+2}
- \delta_{m}
+ \delta_{m-1} +\delta_{m+1} - \delta_{m+2} = 0.
\ee
A convenient way to think of (\ref{funeq1}) is to interpret $\phi$ as the kernel of an operator, according to (\ref{opdef}). 
(\ref{funeq1}) may then be written as the operator equation
$$
(D^3-D^2-D+1)\phi = -D^3+D^2-D+1
$$
which may be easily solved to give
\be\label{opsolfuneq1}
\phi = \frac{1+D^2}{1-D^2}
\ee
as an operator. As a function, $\phi$ is the kernel of the operator\footnote{
Some care must be taken in the interpretation of formula 
(\ref{opsolfuneq1}). It was specified that $m\mapsto\phi_m$ has to be a periodic function. 
This means that the shift operator $D$ has to be viewed here as the periodic shift operator, i.e. 
such that  $D^\nu=Id$. However, in all equations which make use of the formula 
(\ref{opsolfuneq1}) this subtlety is not an issue as the periodic operator will always be applied to 
a periodic sequence, for which there is no difference between the standard shift operator and the 
periodic one.
} 
in (\ref{opsolfuneq1}). What happens now is that in fact for this choice of $\phi$, the Poisson 
bracket $\{\rho_m,\rho_n\}$ also vanishes, so that $\rho$ is a Casimir. This implies the ``first-class 
constraint'' condition in \cite{frs}, although $\rho$ being a Casimir is a stronger condition. 
It is now a straightforward exercise to compute the coefficient involved in the bracket 
$\{\mu_m,\mu_n\}$ of (\ref{murhoPB}). Thus, using the operator presentation of $\phi$, for which 
$2\phi_m - \phi_{m+1}-\phi_{m-1}- \delta_{m+1} + \delta_{m-1}
\sim
D^{-1}[2D-1-D^2]\phi - D + D^{-1}$
$$
\ba
-D^{-1}[(1-D)^2\phi-D^2+1] &= -D^{-1}(1-D)[(1-D)\phi-1-D]\\
&=-D^{-1}(1-D)[(1+D)^{-1}(1+D^2)-1-D]\\
&=
D^{-1}(D-1)(D+1)^{-1}[1+D^2-(1+D)^2]\\
&=2\frac{1-D}{1+D}.
\ea
$$
We obtain the Poisson bracket on $\cC\backslash\cW/G$ in the variable $u=\mu|_{\rho\equiv1}$,
\be\label{redPB1}
\{u_m,u_n\} = \pi_{m-n}u_mu_n + \delta_{m-n+1} - \delta_{m-n-1}
\ee
where $\pi\in Fun(\ZZ/N\ZZ,\RR)$ is the kernel of the operator $(1-D)(D+1)^{-1}$. Here the bracket in (\ref{murho}) has been divided by 2. Alternatively, the Poisson tensor may be written in operator form,
\be\label{redPB2}
P(u)=u\left(\frac{1-D}{1+D}\right)u + D - D^{-1}.
\ee
This is the same as the Poisson tensor obtained in \cite{frs}.

\subsubsection{Choice of $\phi$ adapted to the Toda lattice interpretation}

Let us choose $\phi$ so that the Poisson bracket depends linearly on $\mu$. As the only 
quadratic term in $\mu$ is in the $\{\mu_m,\mu_n\}$ bracket,  we should look for a solution of the 
difference equation
\be\label{funeq2}
2\phi_{m} - \phi_{m+1}-\phi_{m-1} - \delta_{m+1} + \delta_{m-1} = 0.
\ee
In operator form, we have the equation for $\phi$
$$
(D-1)^2\phi + D^2 - 1=0
$$
which has the solution
\be\label{opsolfuneq2}
\phi=\frac{1+D}{1-D}.
\ee
Using this choice for $\phi$ we should now evaluate the coefficients appearing in the other Poisson brackets of (\ref{murhoPB}). In operator form, for the $\{\mu_m,\rho_n\}$ bracket we must compute
$$
\ba
&-D^{-1}\bigl[(D^3-D^2-D+1)\phi + D^3- D^2 + D -1\bigr]\\
&\qquad\qquad\qquad\qquad\qquad\qquad\qquad=
-D^{-1}(D-1)\bigl[(D^2-1)\phi + (D^2+1)\bigr]\\
&\qquad\qquad\qquad\qquad\qquad\qquad\qquad=
D^{-1}(D-1)\bigl[(1+D)^2 - (1+D^2)\bigr]\\
&\qquad\qquad\qquad\qquad\qquad\qquad\qquad= 2(D-1).
\ea
$$
For the $\{\rho_m,\rho_n\}$ bracket we must compute
$$
\ba
D^{-2}\bigl[1-D^4 - (1-D^2)^2\phi\bigr] &= D^{-2}(1-D^2)\bigl[(1+D^2)- (1+D)^2\bigr]\\
&=
-2D^{-1}(1-D^2)\\
&=2(D-D^{-1}).
\ea
$$
Putting these together, we obtain, for the choice of $\phi$ in (\ref{opsolfuneq2}), dividing the right-hand side of (\ref{murhoPB}) by 2,
\be\label{todaPB}
\ba
\{\mu_m,\mu_n\}&=\delta_{m-n+1}\rho_n - \delta_{m-n-1}\rho_m\\
\{\mu_m,\rho_n\}&=(\delta_{m-n+1}-1)\mu_m\rho_n\\
\{\rho_m,\rho_n\}&=(\delta_{m-n+1} - \delta_{m-n-1})\rho_m\rho_n.
\ea
\ee
In operator form, this Poisson structure is given by
\be\label{optodaPB}
\left(\begin{array}{c}\dot\mu \\ \dot\rho\end{array}\right) = 
\left(\begin{array}{cc}D\rho-\rho D^{-1} & \mu(D-1)\rho \\ \rho(1-D^{-1})\mu & \rho(D-D^{-1})\rho\end{array}\right)\cdot
\left(\begin{array}{c}\delta_\mu H \\ \delta_\rho H\end{array}\right)
\ee
which is the ``second Poisson structure'' for the Toda lattice.

It is worth pointing out that the ``lifted version'' of the standard Toda flow to the space $\cW$ 
can be computed as the Hamiltonian flow with Hamiltonian $H(V)=\sum_m\mu_m(V)$. The result 
is
\be
\dot V = D^{-1}(\rho V)\qquad
\hbox{or}
\qquad
\dot V_m = \frac{|V_mV_{m+1}|}{|V_{m-1}V_m|}V_{m-1}.
\ee

\subsubsection{Faddeev-Takhtajan-Volkov obtained as constrained Toda}

\begin{corollary}\label{TodatoFTV}
Applying the constraint $\rho\equiv1$ to the Poisson structure in (\ref{todaPB}, \ref{optodaPB})  
results in the lattice Virasoro structure (\ref{redPB1}, \ref{redPB2}) of \cite{frs}. 
\end{corollary}
\begin{proof}
This follows directly from Proposition \ref{projprop}. A demonstration of the result is given in the Appendix.
\end{proof}

To close this section let us state the result already alluded to and proven indeed 
in \cite{frs},
\begin{proposition}\label{FRStoFTV}
In the variables $S$ on $\cC\backslash\cW/G$, given by the change
of variables $u\mapsto S=uu'$, the Poisson structure (\ref{redPB2}) is represented by
\be\label{opftvPB}
\ba
P(S) &= DS + SD - D^{-1}S-SD^{-1} + SD^{-1}S - SDS\\ 
&\qquad\qquad+ SDS^{-1}DS - SD^{-1}S^{-1}D^{-1}S
\ea
\ee
\end{proposition}

\begin{proof} A demonstration of this result is also given in the Appendix.\end{proof}

\subsubsection{A connection with the Dressing Chain of Veselov and Shabat}

The description of coordinates on $\cC\backslash\cW/G$ can be modified and generalised to incorporate a result from the paper of Veselov and Shabat \cite{vs}. Suppose that $\beta$ is a fixed periodic sequence in $\RR^\times$ and let $B$ be a sequence 
for which $B'=\beta B$ 
(i.e. $B_m=\beta_{m-1}\beta_{m-2}\cdots\beta_0$). For $V$ a quasi-periodic sequence, let $\Gamma(V)\in\cC$ be such that the Wronskian of
$\tilde V=\Gamma V$ has the property, 
$\displaystyle{\frac{w(\tilde V)'}{w(\tilde V)}\equiv\beta}$. We have then 
$\tilde V'' = u\tilde V' - \beta\tilde V$. $u$ is 
an invariant of the $\cC$-action as well as of the $G$ action and so projection from $\cW/G$ to 
$\cC\backslash\cW/G$ is implemented directly in the coordinates $(u,\Gamma)$ by 
forgetting $\Gamma$. The simplest way to implement this is to make the change of variables 
$(u,\Gamma)\mapsto (S,\gamma)=(\beta'^{-1}uu',\Gamma\Gamma')$. This 
results in the expressions
$$
\gamma(V)=\frac{B}{w(V)}\quad\hbox{and}\quad 
S(V)=\frac{|VV''|\,|VV''|'}{|VV'|\,|VV'|''}
$$
for $\gamma$ and $S$, by means of which the Poisson structure on $\cW/G$ is explicitly computed in terms of $(S,\gamma)$. As in the special case $\beta\equiv1$ already discussed, 
the Poisson projection from $\cW/G$ to 
$\cC\backslash\cW/G$ is represented by the algebra $\{S_m,S_n\}$ and by simply ``forgetting'' the contribution of $\gamma$. The formulae in (\ref{redPB1}) and (\ref{redPB2}) are modified in this 
more general case by constraining to $\rho\equiv\beta$ instead of to $\rho\equiv1$.

\begin{note}
This means that the lattice Virasoro algebra made its first appearance, in the more general 
form with $\beta$ an arbitrary periodic sequence, in the article of Veselov and Shabat, although
it seems not to have been recognised as being equivalent to the Faddeev-Takhtajan-Volkov
structure.\footnote{I am grateful to A.Veselov for discussions in which this observation came to light.}
\end{note}

\section{Polygons in $\RR^3$}\label{nu3}

It is instructive to look at the $\nu=3$ case in the same detail.
For $V$ a  sequence in $\RR^3$, for any $m$, the four vectors 
$V_m, V_{m+1}, V_{m+2}, V_{m+3}$ are linearly dependent. Assuming that $V$ is a non-degenerate sequence, the linear 
dependence may be written in the form $V_{m+3}=a_mV_{m+2}-b_mV_{m+1} + \rho_mV_m$. 
In more compact notation, we have
\be\label{lindep3}
V'''=a V'' - b V' + \rho V.
\ee
The sequences $a$, $b$ and $\rho$ depend on the sequence $V$ and they may be expressed explictly:
\be\label{abrho}
w(V)a(V)=|VV'V'''|,\quad w(V)b(V)=|VV''V'''|,\quad w(V)\rho(V)=w(V)'.
\ee
$a$, $b$ and $\rho$ are periodic and $\forall g\in G\ (a(Vg), b(Vg),\rho(Vg))= (a(V),b(V)\rho(V))$. 
Hence, as dim$(\cW)=3N+$dim$(G)$, $(a,b,\rho)$ are good coordinates on $\cW/G$.

Using the expressions in (\ref{abrho}), we may compute the representation of the Poisson
structure on $\cW/G$ in terms of $(a,b,\rho)$ explicitly:
\be\label{abrhoPB}
\left\{\begin{array}{llll}
&\{a_m,a_n\} &= &\bigl(2\phi_{m-n} - \phi_{m-n+1}-\phi_{m-n-1} - \delta_{m-n+1} + \delta_{m-n-1}\bigr)
a_ma_n\\
&{}&{}&\qquad\qquad
+
2\delta_{m-n+1}b_n - 2\delta_{m-n-1}b_m\\
&\{a_m,b_n\} &= &\bigl(\phi_{m-n} +\phi_{m-n+1} - \phi_{m-n+2}-\phi_{m-n-1}\\ 
&{}&{}&\qquad\qquad
- \delta_{m-n} + \delta_{m-n+1} -\delta_{m-n+2} + \delta_{m-n-1}\bigr)a_mb_n\\ 
&{}&{}&\qquad\qquad\qquad\qquad
+
2\delta_{m-n+2}\rho_n - 2\delta_{m-n-1}\rho_m\\
&\{a_m,\rho_n\} &= &\bigl(\phi_{m-n}+\phi_{m-n+2} - \phi_{m-n+3} - \phi_{m-n-1}\\ 
&{}&{}&\qquad\qquad
- \delta_{m-n}+\delta_{m-n+2} - \delta_{m-n+3}+\delta_{m-n-1}\bigr)a_m\rho_n\\
&\{b_m,b_n\} &= &\bigl(2\phi_{m-n} - \phi_{m-n+2}-\phi_{m-n-2} - \delta_{m-n+2} + \delta_{m-n-2}\bigr)
b_mb_n\\
&{}&{}&\qquad\qquad
+
2\delta_{m-n+1}a_m\rho_n - 2\delta_{m-n-1}\rho_ma_n\\
&\{b_m,\rho_n\} &= &\bigl(\phi_{m-n}+\phi_{m-n+1} - \phi_{m-n+3} - \phi_{m-n-2}\\ 
&{}&{}&\qquad\qquad
- \delta_{m-n}+\delta_{m-n+1} - \delta_{m-n+3}+\delta_{m-n-2}\bigr)b_m\rho_n\\
&\{\rho_m,\rho_n\} &= &\bigl(2\phi_{m-n}-\phi_{m-n+3} - \phi_{m-n-3} - \delta_{m-n+3}+\delta_{m-n-3}\bigr)\rho_m\rho_n
\end{array}\right.
\ee

\subsection{Choices for the function $\phi$}
Constraint of the algebra (\ref{abrhoPB}) to $\rho\equiv1$ (or to any fixed periodic sequence
$\rho\equiv\beta$ in $\RR^\times$) is guaranteed to be independent of 
$\phi$, due to Proposition \ref{projprop}. There are three natural special choices for $\phi$ 
which present themselves for closer inspection. The first is the one for which $\rho$ is a Casimir of 
(\ref{abrhoPB}). The second is the one for which the Poisson tensor depends linearly on the 
variable $a$. The third is the one for which the Poisson tensor depends linearly on the 
variable $b$.

\begin{remark} It is not obvious that there exists a choice of $\phi$ for which $\rho$ is a Casimir. 
That this turns out to be possible implies the crucial result in \cite{ss} that it is possible to fix 
their r-matrix in such a way for the constraints to be first-class. In the present context it may be 
found by straightforward computation that imposing say $\{a_m,\rho_n\}=0$ implies both 
$\{b_m,\rho_n\}=0$ and $\{\rho_m,\rho_n\}=0$. Making the resulting choice for $\phi$ is equivalent 
to singling out the r-matrix in \cite{ss}.
\end{remark}

\subsubsection{$\phi$ chosen so that $\rho$ is a Casimir} Let us look for $\phi$ in (\ref{abrhoPB})
to force the brackets $\{a_m,\rho_n\}$ to vanish. The equation for $\phi$ is
$$
(-D^4+D^3+D-1)\phi  = (D^4-D^3+D-1)
$$
which has the solution
\be\label{rhocas3}
\phi = \frac{1+D^3}{1-D^3}
\ee
The following proposition may be checked by substitution of (\ref{rhocas3}) in (\ref{abrhoPB}).
\begin{proposition}
When the function $\phi$ is the kernel of the operator in (\ref{rhocas3}) $\rho$ is a Casimir function of the Poisson structure defined by (\ref{abrhoPB}).
\end{proposition}

It follows that the reduced Poisson structure on $\cC\backslash\cW/G$ is given in local 
coordinates $(a,b)$, with $\rho\equiv1$, by the relevant brackets in (\ref{abrhoPB}) in which 
$\phi$ is the operator in 
(\ref{rhocas3}):
\be\label{rhocas3PB}
\ba
&\left(\begin{array}{c}\dot a \\ \dot b\end{array}\right)
= P_0(a,b)\cdot \left(\begin{array}{c}\delta_aH \\ \delta_bH\end{array}\right),\\
&\hbox{with}\\
&P_0(a,b)=\left(\begin{array}{cc}
{\begin{array}{ll}&a(1+D+D^2)^{-1}(1-D^2)a\\ &\qquad\qquad+ Db - bD^{-1}\end{array}} 
& 
{\begin{array}{ll}&a(1+D+D^2)^{-1}(D-D^2)b\\ &\qquad\qquad + D^2-D^{-1}\end{array}}
\\ \\ \\
\begin{array}{ll}&b(1+D+D^2)^{-1}(1-D)a\\ &\qquad\qquad + D-D^{-2}\end{array}
 & 
 \begin{array}{ll}&b(1+D+D^2)^{-1}(1-D^2)b\\ &\qquad\qquad+ aD - D^{-1}a\end{array}
\end{array}\right).
\ea
\ee

\subsubsection{$\phi$ chosen so that the Poisson tensor depends linearly on $a$} Choosing 
\be\label{noquada}
\phi= \frac{1+D}{1-D}
\ee
solves the equation
$$
(2-D-D^{-1})\phi-D+D^{-1}=0
$$
and forces the dependence on $a$ to be linear. The Poisson bracket (\ref{abrho}) with $\phi$ substituted from (\ref{noquada}) is
\be\label{noquadaPB}
\ba
&\left(\begin{array}{c}\dot a \\ \dot b \\ \dot\rho\end{array}\right)
= P_1(a,b,\rho)\cdot\left(\begin{array}{c}\delta_aH \\ \delta_bH\\ \delta_\rho H\end{array}\right)\\
&\hbox{with}\ \ P_1(a,b,\rho)=\\
&\left(\begin{array}{ccc}Db-bD^{-1} & 
\begin{array}{ll}&a(D-1)b\\ &+ D^2\rho-\rho D^{-1}\end{array} 
& a(1-D^2)\rho \\ \\
\begin{array}{ll}&b(1-D^{-1})a\\ & +D\rho-\rho D^{-2}\end{array} & 
\begin{array}{ll}&b(D-D^{-1})b\\ &+aD\rho-\rho D^{-1}a\end{array} & b(D-D^{-1})(D+1)\rho \\ \\
\rho(D^{-2}-1)a & \rho(1+D^{-1})(D-D^{-1})b &\quad \rho(D^2+D-D^{-1}-D^{-2})\rho\end{array}\right).
\ea
\ee

\subsubsection{$\phi$ chosen so that the Poisson tensor depends linearly on $b$} Choosing 
\be\label{noquadb}
\phi= \frac{1+D^2}{1-D^2}
\ee
solves the equation
$$
(1+D-D^2-D^{-1})\phi-1+D-D^2+D^{-1}=0
$$
and forces the dependence on $b$ to be linear. The Poisson bracket (\ref{abrho}) with $\phi$ substituted from (\ref{noquadb}) is
\be\label{noquadabPB}
\ba
&\left(\begin{array}{c}\dot a \\ \dot b \\ \dot\rho\end{array}\right)
= P_2(a,b,\rho)\cdot\left(\begin{array}{c}\delta_aH \\ \delta_bH\\ \delta_\rho H\end{array}\right)\\
&\hbox{with}\ \ P_2(a,b,\rho)=\\
&\left(\begin{array}{ccc}
a\displaystyle{\frac{1-D}{1+D}}a+Db-bD^{-1} \quad&  D^2\rho-\rho D^{-1} & 
aD\displaystyle{\frac{D-1}{D+1}}\rho \\ \\
 D\rho-\rho D^{-2}& aD\rho-\rho D^{-1}a & b(D-1)\rho \\ \\
 \rho D^{-1}\displaystyle{\frac{D-1}{1+D}}a & \rho(1-D^{-1})b 
 & \quad\rho\displaystyle{\frac{D-1}{D+1}}(D+1+D^{-1})\rho\end{array}\right).
\ea
\ee

\subsection{Discussion of $\nu=3$ examples} 

The Poisson structures presented in equations (\ref{noquadaPB}) and in (\ref{noquadabPB}) 
are the same as the ones related to different ``\emph{extended Toda}'' systems, 
see for example \cite{car}. This means that by seeing those structures in the context of the 
present article, they are clearly compatible with one another, as their sum is the special case 
of (\ref{abrho}) for which $\phi$ is the sum of the kernels of the operators in (\ref{noquada}) 
and (\ref{noquadb}), which is just some other odd function.

On the other hand, the compatibility of the two Poisson structures (\ref{noquadaPB}) and 
(\ref{noquadabPB}) in the usual context of extended Toda systems seems not only surprising, but
totally inappropriate as it appears to amount to the comparison of two structures defined on 
different spaces: that is, $P_1$ is naturally defined on the set of difference operators of the form 
$\{D+u_0+u_{-1}D^{-1}+u_{-2}D^{-2}\}$ and $P_2$ is naturally defined on the set of difference 
operators of the form $\{D^2+u_1D+u_0+u_{-1}D^{-1}\}$. In the setting of the present article these two spaces \emph{are} just the same, but in the standard setting of Toda systems they 
are not.

The Poisson structures $P_1$ and $P_2$ in (\ref{noquadaPB}) and in (\ref{noquadabPB})
were obtained in \cite{car} by applying a formula in \cite{lp,or} for which the two extended Toda 
spaces are not treated in the same way at all.

\section{A result for the general $\nu$ case}\label{gennu}

We observe that the different choices of $\phi$ of especial interest have a natural 
form. That is, for \(\nu=2\) we make the choices
\[
\phi=\frac{1+D^2}{1-D^2}\quad\hbox{and}\quad\phi=\frac{1+D}{1-D}
\]
and for \(\nu=3\) the special choices are
\[
\phi=\frac{1+D^3}{1-D^3},\quad 
\phi=\frac{1+D^2}{1-D^2},\quad\hbox{and}\quad
\phi=\frac{1+D}{1-D}.
\]
We may imagine that the obvious pattern shows itself in the case of general $\nu$. First of all
we need to draw attention to a particularly useful feature of the choices above. In each case, 
the first choice is the one which makes the reduction to the space $\cC\backslash\cW/G$ explicit,
whilst the others have the property that they render the Poisson 
structure on $\cW/G$ linear in one or other of the variables. This property of linearity is
important in the context of bi-hamiltonian systems as it guarantees the possibility to find a vector 
field satisfying the conditions of Proposition \ref{GF} and so to deform the Poisson structure in 
such a way to obtain another one which is necessarily compatible with it. We may note that not 
only does this observation apply to the formulae (\ref{optodaPB}, \ref{noquadaPB}, 
\ref{noquadabPB}), but it corresponds to standard known results for Toda and extended Toda 
systems, see for example \cite{car}.

Let us look at what happens in the general case. First of all, observing that the $\nu+1$ 
vectors $\RR^\nu\owns V,V',V'',\dots, V^{(\nu)}$ obey some linear dependence relation and that
the nondegeneracy condition $w(V)\neq0$ allows us to assume the coefficient of 
$V^{(\nu)}$ in this relation to be $1$, we may write
\be\label{lindepgen}
V^{(\nu)} - a^{(\nu-1)}V^{(\nu-1} + \cdots + (-1)^{(\nu-r)}a^{(\nu-r)}V^{(\nu-r)} + \cdots + (-1)^\nu a^{(0)}V=0.
\ee
Hence the quasi-periodic sequence $V\in\cW$ engenders the element 
$L=D^\nu - a^{(\nu-1)}D^{\nu-1}+\cdots + (-1)^{\nu-1}a^{(1)}D+(-1)^\nu a^{(0)}\in{\mathcal S}_\nu$ and
the fields $(a^{(0)},a^{(1)},\dots,a^{(\nu-1)})$ are a good set of coordinates on $\cW/G$. Let us
set ${\mathbf a}= (a^{(0)},a^{(1)},\dots,a^{(\nu-1)})$. 
For $k=1,\dots,\nu-1$, define the functions 
$\alpha^{(k)}$ on $\cW$ by
\[
\alpha^{(k)} = |V \dots V^{(\nu)}|\overset{\overset{k}{\vee}}{\phantom{e}},\ \hbox{i.e.}\ 
\alpha^{(k)}_m=|V_m\dots V_{m+k-1}V_{m+k+1}\dots V_{m+\nu}|
\]
where the symbol $\overset{k}{\vee}$ on the determinant means that the $k$th term is not present, as is clear in the second expression. It is easy to check that 
\be\label{defas}
\hbox{For } 1\leq k\leq\nu-1,\qquad a^{(k)}=\frac{\alpha^{(k)}}{w}\qquad\hbox{and }\qquad a^{(0)}=\frac{w'}{w}.
\ee
For a general value of $\nu$ a generalisation of the results of Sections \ref{nu2} 
and \ref{nu3} provides positive answers to the questions
\begin{quote}
\emph
{
For $k\in\{1,\dots,\nu-1\}$, is there a choice of $\phi$ such that the Poisson tensor 
$P_\phi({\mathbf a})$ on $\cW/G$, when represented in the coordinates 
$\mathbf a$, is linear in the coordinate $a^{(k)}$: so that
\(\displaystyle{\frac{d^2}{dt^2}P(a^{(0)},\dots, a^{(k)}+t,\dots, a^{(\nu-1)})=0}\)?
}
\end{quote}
\begin{quote}
\emph{
Is there a choice of $\phi$ such that $a^{(0)}$ is a Casimir?}
\end{quote}
in the form of the following theorem, whose proof is given in Appendix B.

\begin{theorem}\label{phichoose}
For the family of $\phi-$dependent Poisson brackets on ${\mathcal S}_\nu=\cW/G$, with respect 
to the natural coordinates $\mathbf a$ on ${\mathcal S}_\nu$, given by
\({\mathcal S}_\nu \owns 
L=D^\nu - a^{(\nu-1)}D^{\nu-1}+\cdots + (-1)^{\nu-1}a^{(1)}D+(-1)^\nu a^{(0)}\)\\
(i) the choice
\[
\phi^{(0)}:=\frac{1+D^\nu}{1-D^\nu}
\]
forces $a^{(0)}$ to be a Casimir; \\
(ii) for $0<k<\nu$, the choice
\[
\phi^{(k)}:=\frac{1+D^{\nu-k}}{1-D^{\nu-k}}
\]
forces the Poisson bracket to have linear dependence on $a^{(k)}$.
\end{theorem}

The following corollary tells us that we can introduce a spectral parameter in a straightforward way 
for any of the choices $\phi=\phi^{(k)}$ for $0\leq k\leq\nu$

\begin{corollary}
A spectral parameter $\lambda$ can be introduced as follows: With the choice
$\phi=\phi^{(0)}$ and $\beta$ any fixed periodic sequence, $\rho\equiv\beta$ can be replaced by $\rho\equiv\beta+\lambda$. With the choice
$\phi=\phi^{(k)}$, for $0<k<\nu$, $a^{(k)}$ can be replaced by $a^{(k)}+\lambda$.
\end{corollary}

\section{Conclusion and perspectives}

As was shown in \cite{ms}, introducing a Poisson structure on the space $\cW$ with 
the property that the actions of $\cC$ and $G$ on $\cW$ be Poisson is a natural setting for 
the Poisson description of the space ${\mathcal S}_\nu$, by the identification 
${\mathcal S}_\nu=\cW/G$. In fact we have proceeded on the basis of an initial \emph{family} 
of (compatible) Poisson structures on $\cW$ parametrised by an arbitrary odd function
$\phi$ appearing in the formula for the original Poisson bracket on $\cW$. Consequently there
is a family of Poisson structures on ${\mathcal S}_\nu=\cW/G$, still parametrised by the 
arbitrary odd function $\phi$. It has been shown however that the Poisson structure on 
$\cC\backslash\cW$, obtained by reducing the one on $\cW$, is independent of $\phi$.
It follows that the projection $\Pi:\cW/G\rightarrow\cC\backslash\cW/G$ takes
the $\phi$-dependent family of Poisson structures to a single $\phi$-independent structure.
The map $\Pi$ is represented in the space ${\mathcal S}_\nu$ by imposing the 
constant-Wronskian constraint, $u_0\equiv(-1)^\nu$, where 
${\mathcal S}_\nu=\{L=D^\nu+u_{\nu-1}D^{\nu-1}+\cdots +u_1D+u_0\}$ and therefore the Poisson 
structure can be discovered by applying Dirac's method of constraints. It has been shown how the 
freedom in the choice of $\phi$ for the result of applying the projection $\Pi$ may be used to 
simplify the computations of the reduction. In the cases $\nu=2,3$ it has been noticed that other 
special choices for the function $\phi$ lead to the (\emph{second}) Poisson structures for 
Toda and extended Toda systems. A simple corollary is that a constraint of the second Toda
Poisson bracket yields the lattice Virasoro Poisson bracket and that the same thing happens
for the extended Toda systems and the higher rank versions of lattice Virasoro, known as 
lattice W-algebras.

It seems undeniable that the point of view presented first in \cite{ms} and here developed further 
cannot fail to be a useful one, extending as it manifestly does, the $sl_\nu$ cases of results presented in 
\cite{frs} and \cite{ss}. It will be interesting to discover how to extend the approach taken here
to encapsulate all of the cases in \cite{ss} covering as they do all the classical root 
systems and not just the ones of type $A_n$. It is expected that such a generalisation may 
have useful applications for the construction of examples of interesting new Frobenius manifolds, 
\cite{d}: see for example the articles \cite{cdz, dz}.

Perhaps the most interesting outcome of the analysis developed in the present article is the 
understanding of how a spectral parameter can be introduced. It is expected that this could 
lead to a constructive method for generating integrable flows. It would be interesting to 
understand how the simple dependence of the fields $a^{(k)}$ on the spectral parameter $\lambda$
for different choices of $\phi$ are reflected in some more complicated dependence on
$\lambda$ of the wave-functions $V\in\cW$.

A tantalising problem of some interest to the author is to discover the link between a result of 
the article \cite{ost} and the Poisson structure on $\cW$. It seems inconceivable that such 
a link should not exist, but so far it has not proved possible to identify one. The original hope was 
that the Poisson structure found in \cite{ost} might prove to be compatible with the one on 
$\cC\backslash\cW/G$, but this is not the case. This link is especially interesting because the 
integrable system described in \cite{ost} does have a Lax representation depending on a 
spectral parameter.

The Poisson structure presented in the formula (\ref{PBonW}) and in
Proposition \ref{basic} is not an original one. It is a representation of the so-called 
``\emph{exchange algebra}''. The quantum version of the exchange algebra, from which the 
Poisson case can be extracted as a classical limit, is explained in the lecture notes \cite{f} 
of Faddeev. It appeared already in the article \cite{stsrims} and has been made use of in many
contexts since then, see for example \cite{bfpmontreal}. More notably it was also used in a 
form almost the same as that exploited here, by Babelon, especially in \cite{Bab}. Indeed 
the only thing added here to Babelon's work is the recognition of the role of Poisson Lie groups 
and of the additional function $\phi$. 

\section{Appendix A: missing proofs}

\subsection{Proof of Proposition \ref{momprop}} 

The claim was that $w$ defines a momentum map for the action of $\cC$ on $\cW$. The following lemmas are used in the proof.
\begin{lemma}\label{siglem}
$\sigma_{m+k} = \sigma_m +\delta_m+2[\delta_{m+1}+\cdots+\delta_{m+k-1}] + \delta_{m+k}$. 
\end{lemma}
\begin{proof}
The case $k=1$, $\sigma_{m+1}-\sigma_m=\delta_m+\delta_{m+1}$ is easy to check. The general case may then be proved by induction. Equivalently, the operator whose kernel is the function $m\mapsto\sigma_{m}$ is the Cayley transform of $D$,
\[
\sigma = \frac{D+1}{D-1}
\]
from which the same formula is found directly.
\end{proof}

\begin{lemma}\label{actionlem}
Let $\xi_1,\xi_2\dots,\xi_\nu\in\RR^\nu$ and $X\in\g$. 
Then
$\sum_{k=1}^\nu|\xi_1\dots(X\xi_k)\dots\xi_\nu|=0.$
\end{lemma}
\begin{proof}
For $g(t)=\exp(tX)$, writing $\tilde\xi_k(t)=g(t)\xi_k\ \forall k$,
\[
\ba
\sum_{k=1}^\nu|\xi_1\dots\xi_{k-1}(X\xi_k)\xi_{k+1}\dots\xi_\nu|&=
\left.\frac{d}{dt}\right|_{t=0}|\tilde\xi_1(t)\dots\tilde\xi_k(t)\dots\tilde\xi_\nu(t)|\\
&=\left.\frac{d}{dt}\right|_{t=0}|g(t)|\,|\xi_1\dots\xi_\nu|=0.
\ea
\]
\end{proof}

\begin{proofof}{Proposition}{momprop}

Writing $R=\sum_iX_i\otimes Y_i$ and $C=\sum_jP_j\otimes Q_j$ for suitable $X_i,Y_i,P_j,Q_j\in\g$, 
we may compute
\[
\ba
\{w_m,V_n\}&=
\sum_{k=0}^{\nu-1}\Bigl(\sum_i |V_m\dots V_{m+k-1}(V_{m+k}X_i)V_{m+k+1}
\dots V^{(\nu-1)}| V_nY_i\\
&\qquad\quad-\sigma_{m+k-n}|V_m\dots V_{m+k-1}V_nV_{m+k+1}\dots V^{(\nu-1)}| V_{m+k} \\
&\qquad\qquad+\phi_{m+k-n}|V_m\dots V_{m+k}\dots V^{(\nu-1)}| V_n\Bigr)\\
&=
\sum_i\Bigl(\sum_{k=0}^{\nu-1}|V_m\dots (V_{m+k}X_i)\dots V^{(\nu-1)}|\Bigr)V_nY_i\\
&\qquad-
\sum_{k=0}^{\nu-1}\sigma_{m+k-n}|V_m\dots V_{m+k-1}V_nV_{m+k+1}\dots V^{(\nu-1)}|V_{m+k}\\
&\qquad\qquad+\Bigl(\sum_{k=0}^{\nu-1}\phi_{m+k-n}\Bigr)w_mV_n,
\ea
\]
where use has been made of the property (\ref{Cexprop}). The first term is zero by Lemma \ref{actionlem}. Using Lemma \ref{siglem}, the general term 
in the second sum is 
$\bigl(\sigma_{m-n} +\delta_{m-n} + 2[\delta_{m-n+1}+\cdots+\delta_{m-n+k-1}] + \delta_{m-n+k}\bigr)|V_m\dots V_{m+k-1}V_nV_{m+k+1}\dots V^{(\nu-1)}|$. Hence
\[
\ba
&\sigma_{m-n+k}|V_m\dots V_n\dots V^{(\nu-1)}|V_{m+k}
=
(\sigma_{m-n}+\delta_{m+k-n})|V_m\dots V_n\dots V^{(\nu-1)}|V_{m+k}\\ 
&\qquad\qquad
= \sigma_{m-n}|V_m\dots V_n\dots V^{(\nu-1)}|V_{m+k} + \delta_{m+k-n}w_mV_n\\
&\qquad\qquad=
\sigma_{m-n}\sum_j|V_m\dots (V_{m+k}P_j)\dots V^{(\nu-1)}|V_nQ_j + (\sigma_{m-n}+\delta_{m+k-n})w_mV_n
\ea
\]
and, summing over $k$ and using Lemma \ref{actionlem}, we obtain
\[
\{w_m,V_n\}=\left(\sigma_{n-m}+\sum_{r=0}^{\nu-1}\phi_{m+r-n}
+\sum_{r=1}^{\nu-1}\delta_{m+r-n}\right)w_mV_n.
\]
\end{proofof}

\subsection{Proof of Proposition \ref{projprop}}
Here the notation $\langle\ \,,\ \rangle$ is used to denote the standard scalar product both on 
$\RR^\nu$ and on $\RR^{\nu-1}$.
The following Lemma is used for the proof of the Proposition.
\begin{lemma}
Let $X\in\g$. Then, in terms of local coordinates $v$ on $\RR\PP^{\nu-1}$, 
the projective action $X\cdot v$ of $\g$ on $\RR\PP^{\nu-1}$ 
is determined by
$\langle\alpha, X\cdot v\rangle = \langle(\alpha,-\langle\alpha,v\rangle),(v,1)X\rangle\\
\ \ \forall\alpha\in\RR^{\nu-1}$. 

\end{lemma}
\begin{proof}
To see this, we projectivise the action of $\exp tX\in G$ on $\RR^\nu$ and differentiate with respect to $t$. Thus, writing 
$\displaystyle{X=\left(\begin{array}{cc}A & b^T \\c & d\end{array}\right)}$ 
(where $c$ and $b$ are \emph{row} vectors), we get
$$
\ba
X\cdot v &= \left.\frac{d}{dt}\right|_{t=0}\frac{1}{1+td + tvb^T}(v+tvA+tc)\\
&=
vA+c-dv-(vb^T)v
\ea
$$
It is now straightforward to check that
$$
\langle\alpha,X\cdot v\rangle = \langle(\alpha, -\langle\alpha,v\rangle),(v,1)X\rangle. 
$$
\end{proof}

\begin{proofof}{Proposition}{projprop} For $\alpha$ and $\beta$ some pair of sequences in 
$\RR^{\nu-1}$, let $F,H\in\cinf(\cW)^\cC$, with $F(\chi(v,1))=\langle\alpha,v\rangle$ and 
$H(\chi(v,1))=\langle\beta,v\rangle$. Then 
$dF(v,1)=(\alpha,-\langle\alpha,v))$ and $dH(v,1)=(\beta,-\langle\beta,v\rangle)$.
Writing, as before, $R=\sum_iX_i\otimes Y_i$ and making use of the Lemma, 
gives
$$
\ba
\{F,H\}(v,1)&=
\sum_{m,n}\Bigl\langle(\alpha,-\langle\alpha,v\rangle)_m\otimes
(\beta,-\langle\beta,v\rangle)_n, (v_m,1)
\otimes(v_n,1)\\
&\qquad\qquad\qquad\qquad\qquad\qquad[R+\sigma_{m-n}(C+Id\otimes Id)+\phi_{m-n}]\Bigr\rangle\\
&=
\sum_{m,n}\sum_i
\langle\alpha_m,X_i\cdot v_m\rangle\langle\beta_n,Y_i\cdot v_n\rangle\\
&\qquad+
\sum_{m,n}\sigma_{m-n}(v_m^T,1)
\left(\begin{array}{c}\beta_n \\-\langle\beta_n,v_n\rangle\end{array}\right)
(v_n^T,1)\left(\begin{array}{c}\alpha_m \\-\langle\alpha_m,v_m\rangle\end{array}\right)\\
&\qquad\qquad+\sum_{m,n}\phi_{m-n}(\langle\alpha_m,v_m\rangle-\langle\alpha_m,v_m\rangle)
(\langle\beta_n,v_n\rangle-\langle\beta_n,v_n\rangle)\\
&=
\sum_{m,n}\langle\alpha_m\otimes\beta_n, (v_m\otimes v_n)\cdot R\rangle
+ \sum_{m,n}\sigma_{m-n}\langle\alpha_m\otimes\beta_n, (v_n-v_m)\otimes(v_m-v_n)\rangle\\
&=
\sum_{m,n}\langle\alpha_m\otimes\beta_n, (v_m\otimes v_n)\cdot R - \sigma_{m-n}(v_m-v_n)\otimes(v_m-v_n)\rangle
\ea
$$
\end{proofof}

\subsection{Proof of Corollary \ref{TodatoFTV}}
The point of this Corollary is that indeed it follows directly from Proposition \ref{projprop}. However it is straightforward to check the claim directly by using the Poisson tensors in their operator 
forms. The constraint $\rho\equiv1$ is a second-class constraint for the Poisson structure in 
(\ref{optodaPB}) (i.e. $\det(\{\rho_m,\rho_n\}_{\rho\equiv1})\neq0$) and we may apply the 
Dirac formula for a second-class constraint:
\emph{If in local coordinates $(x,y)$ on a Poisson manifold 
($x\in\RR^k, y\in\RR^l$ say), the constraint is supposed to be to the submanifold given by fixing 
the $y$-coordinates $y=y_0$, then if the Poisson structure has the form
\[
\left(\begin{array}{c}\dot x \\ \dot y\end{array}\right) = 
\left(\begin{array}{cc}A & B \\ -B^* & C\end{array}\right)\cdot
\left(\begin{array}{c}\delta_x H \\ \delta_y H\end{array}\right)
\]
with $\det(C|_{y=y_0})\neq0$,
the constrained Poisson structure is given in local coordinates $x\in\RR^k$ by
\[
\dot x = \bigl[A + BC^{-1}B^*\bigr]_{y=y_0}\cdot\delta_xH
\]
}
Applying this formula to (\ref{optodaPB}) with 
$A=D\rho-\rho D^{-1}, B=\mu(D-1)\rho, C=\rho(D-D^{-1})\rho$, yields the constraint of the second Toda Poisson structure to $\rho\equiv1$:
\[
\ba
\dot u =(\dot \mu|_{\rho\equiv1}) &= 
\Bigl[D-D^{-1} - u(D-1)(D-D^{-1})^{-1}(1-D^{-1})u\Bigr]\delta_uH\\
&=
\Bigl[D-D^{-1} - u(D-1)(D^2-1)^{-1}(D-1)u\Bigr]\delta_uH\\
&=\left[u\left(\frac{1-D}{1+D}\right)u +D - D^{-1}\right]\delta_uH.
\ea
\]
\hfill\(\square\)

Generalisation of this proof to the constraint $\rho\equiv\beta$, for some fixed periodic sequence
$\beta\in Fun(\ZZ/N\ZZ,\RR^\times)$, is straightforward.

\subsection{Proof of Proposition \ref{FRStoFTV}}

For $H\in \cinf(Fun(\ZZ/N\ZZ,\RR))$, let $X=\delta_uH$, i.e. 
\[
\forall\Delta\in Fun(\ZZ/N\ZZ,\RR)\ \
\left.\frac{d}{dt}\right|_{t=0}H(u+t\Delta)=\langle\Delta,\delta_uH\rangle = \sum_{n=0}^{N-1}\Delta_n(\delta_uH)_n\
=\sum_{n=0}^{N-1}\Delta_nX_n.
\]
Then the Hamiltonian vector field $\XX_H$ is
\[
\dot u = \XX_H(u) = \left( u\left[\frac{1-D}{1+D}\right]u + D - D^{-1}\right)X.
\]
Now consider $H(u)=F(uu')=F(S)\Rightarrow X= u'\delta_SF + D^{-1}(u\delta_SF)$. Thus
\[
\ba
\dot S = \XX_H(S)&= \dot uu'+u\dot u'=(u'+uD)\dot u\\ 
&= 
(u'+uD)\left( u\left[\frac{1-D}{1+D}\right]u + D - D^{-1}\right)(u'+D^{-1}u)\delta_SF.
\ea
\]
This simplifies to give the result
\[
\ba
\dot S &= \Bigl(DS + SD - D^{-1}S-SD^{-1} + SD^{-1}S - SDS\\ 
&\qquad\qquad+ SDS^{-1}DS - SD^{-1}S^{-1}D^{-1}S\Bigr)\delta_SF
\ea
\]
as promised.
\hfill\(\square\)

Again, generalisation to the constraint $\rho\equiv\beta$ is straightforward.

\section{Appendix B: proofs for Section \ref{gennu}}

Before beginning let us recall that the symbol $\overset{k}{\vee}$
on a determinant signifies that the 
$k$th term is missing. Usually this symbol will be placed at the end, but sometimes it will sit at a 
suitable location in the middle of the determinant. We have
\[
\ba
\{\alpha^{(k)}_m,V_n\}&=\{|V_m\dots V_{m+k-1}V_{m+k+1}\dots V_{m+\nu}|,V_n\}\\
&=
\left(\sum_{r=0,r\neq k}^\nu\phi_{m-n+r}\right)\alpha^{(k)}_mV_n 
+ \sigma_{m-n}|V_nV_{m+1}\dots V_{m+\nu}|\overset{\overset{k}{\vee}}{\phantom{e}}V_m\\
+\ &\sigma_{m-n+1}|V_mV_n\dots V_{m+\nu}|\overset{\overset{k}{\vee}}{\phantom{e}}V_{m+1}
+\cdots+
\sigma_{m-n+k-1}|V_m\dots V_nV_{m+k+1}\dots V_{m+\nu}|V_{m+k-1}\\
&
+\sigma_{m-n+k+1}|V_m\dots V_{m+k-1}V_n\dots V_{m+\nu}|V_{m+k+1}
+\cdots+
\sigma_{m-n+\nu}|V_m\dots V_n|\overset{\overset{k}{\vee}}{\phantom{e}}V_{m+\nu},
\ea
\]
For $0\leq r\leq k-1$, let us use Lemma \ref{siglem} and consider the contribution of
\[
\ba
&
\sigma_{m-n+r}|V_m\dots V_n\dots V_{m+\nu}|\overset{\overset{k}{\vee}}{\phantom{e}}V_{m+r}\\
&\qquad=
\bigl(\sigma_{m-n}+\delta_{m-n}+2(\delta_{m-n+1}\cdots+\delta_{m-n+r-1})+\delta_{m-n+r}\bigr)
|V_m\dots V_n\dots V_{m+\nu}|\overset{\overset{k}{\vee}}{\phantom{e}}V_{m+r}\\
&\qquad=
\bigl(\sigma_{m-n}+\delta_{m-n+r}\bigr) 
|V_m\dots V_n\dots V_{m+\nu}|\overset{\overset{k}{\vee}}{\phantom{e}}V_{m+r}.
\ea
\]
For $k+1\leq r\leq\nu$, again using Lemma \ref{siglem}, consider now the contribution of
\[
\ba
&
\sigma_{m-n+r}|V_m\dots V_{m+k-1}V_{m+k+1}\dots V_n\dots V_{m+\nu}|V_{m+r}=\\
&
\bigl(\sigma_{m-n}+\delta_{m-n}+2(\delta_{m-n+1}\dots+\delta_{m-n+r-1})+\delta_{m-n+r}\bigr)
|V_m\dots V_{m+k-1}V_{m+k+1}\dots V_n\dots V_{m+\nu}|V_{m+r}\\
&\quad=
\bigl(\sigma_{m-n}+2\delta_{m-n+k}+\delta_{m-n+r}\bigr) |V_m\dots V_{m+k-1}V_{m+k+1}\dots V_n\dots V_{m+\nu}|V_{m+r}\\
&\quad=
\sigma_{m-n}|V_m\dots V_n\dots V_{m+\nu}|\overset{\overset{k}{\vee}}{\phantom{e}}V_{m+r} 
+
\delta_{m-n+r}|V_m\dots V_{m+\nu}|\overset{\overset{k}{\vee}}{\phantom{e}}V_n\\
&\qquad\quad
+ 
2(-1)^{r-k-1}\delta_{m-n+k}|V_m\dots V_{m+\nu}|\overset{\overset{r}{\vee}}{\phantom{e}}V_{m+r}\\
&\quad=
\sigma_{m-n}|V_m\dots V_n\dots V_{m+\nu}|\overset{\overset{k}{\vee}}{\phantom{e}}V_{m+r} \\
&\qquad\qquad
+
\delta_{m-n+r}\alpha^{(k)}_mV_n - 2(-1)^{r-k}\delta_{m-n+k}\alpha^{(r)}_mV_{m+r}.
\ea
\]
Putting all of these together and using Lemma \ref{actionlem} in the same way it was used in the
proof of Proposition \ref{momprop}, we obtain
\be\label{avpb1}
\ba
\{\alpha^{(k)}_m,V_n\}&=
\left(\sigma_{m-n} 
+{\sum_{r=0}^\nu}{}_{r\neq k} \phi_{m-n+r}+\sum_{r=1}^\nu{}_{r\neq k}\delta_{m-n+r} \right)\alpha^{(k)}_mV_n\\
&\quad
-
2(-1)^{\nu-k}\delta_{m-n+k}w_m\left(V_{m+\nu} 
+
\sum_{r=1}^{\nu-k-1}(-1)^{\nu-k-r}a^{(k+r)}_mV_{m+k+r}\right)\\
&=
\left(\sigma_{m-n} 
+\sum_{r=0}^\nu{}_{r\neq k}\phi_{m-n+r}+\sum_{r=1}^\nu{}_{r\neq k}\delta_{m-n+r} \right)\alpha^{(k)}_mV_n\\
&\qquad
-
2(-1)^{\nu-k}\delta_{m-n+k}w_mX_{m,n}\qquad\qquad\hbox{say},\\
\hbox{with}\quad
X_{m,n}&=
V_{n+\nu-k} + \sum_{r=1}^{\nu-k-1}(-1)^{\nu-k-r}a^{(k+r)}_mV_{n+r}.
\ea
\ee
Using the relation (\ref{lindepgen}) this equation may also be written in the alternative form
\be\label{avpb2}
\ba
\{\alpha^{(k)}_m,V_n\}&=
\left(\sigma_{m-n} 
+\sum_{r=0}^\nu{}_{r\neq k}\phi_{m-n+r}+\sum_{r=1}^\nu{}_{r\neq k}\delta_{m-n+r}
\right)\alpha^{(k)}_mV_n\\
&\quad
+
2(-1)^{\nu-k}\delta_{m-n+k}w_m\left(
\sum_{r=0}^k(-1)^{\nu-k+r}a^{(k-r)}_mV_{m+k-r}\right)\\
&=
\left(\sigma_{m-n} 
+\sum_{r=0}^\nu{}_{r\neq k}\phi_{m-n+r}+\sum_{r=1}^\nu{}_{r\neq k}\delta_{m-n+r} \right)\alpha^{(k)}_mV_n\\
&\qquad
+
2\delta_{m-n+k}w_mY_{m,n}\qquad\qquad\hbox{say},\\
\hbox{with}\quad
Y_{m,n}&=
\sum_{r=0}^k(-1)^{r}a^{(k-r)}_mV_{n-r}.
\ea
\ee
A much simpler computation leads in a similar fashion to
\be\label{wvpb}
\{w_m,V_n\}=\left(\sigma_{m-n}+\sum_{r=0}^{\nu-1}\phi_{m-n+r}
+\sum_{r=1}^{\nu-1}\delta_{m-n+r} \right)w_mV_n.
\ee
Using (\ref{wvpb}) we may compute
\be\label{wwpb}
\{w_m,w_n\}=\sum_{l=0}^{\nu-1}
\left(\sigma_{m-n-l}+\sum_{r=0}^{\nu-1}\phi_{m-n+r-l}+\sum_{r=1}^{\nu-1}\ 
\delta_{m-n+r-l} \right)w_mw_n
\ee
and
\be\label{walkpb}
\{w_m,\alpha^{(k)}_n\}
=
\sum_{l=0}^{\nu}{}_{r\neq k}
\left(\sigma_{m-n-l}+\sum_{r=0}^{\nu-1}\phi_{m-n+r-l}
+\sum_{r=1}^{\nu-1}\delta_{m-n+r-l} \right)w_m\alpha^{(k)}_n,
\ee
which implies
\be\label{alkwpb}
\{\alpha^{(k)}_m, w_n\}
=
\sum_{l=0}^{\nu}{}_{r\neq k}
\left(\sigma_{m-n+l}+\sum_{r=0}^{\nu-1}\phi_{m-n-r+l}
-\sum_{r=1}^{\nu-1}\delta_{m-n-r+l} \right)\alpha^{(k)}_m w_n.
\ee
Now, using (\ref{avpb1}) and (\ref{avpb2}), let us suppose that
$k< j$ and compute 
\be\label{kjpb}
\ba
\{\alpha^{(k)}_m,\alpha^{(j)}_n\}&=
\sum_{l=0}^\nu{}_{l\neq j}\left(\sigma_{m-n-l}
+\sum_{r=0}^\nu{}_{r\neq k}\ \phi_{m-n+r-l}
+\sum_{r=1}^\nu{}_{r\neq k}\ \delta_{m-n+r-l}\right)\alpha^{(k)}_m\alpha^{(j)}_n\\
&
-
2(-1)^{\nu-k}w_m\Bigl(\delta_{m-n+k}|X_{m,n}V_{n+1}.\!
\overset{\overset{j}{\vee}}{\phantom{e}}\!.\,. V_{n+\nu}|\\
&+
\delta_{m-n+k-1}|V_nX_{m,n+1}.\!
\overset{\overset{j}{\vee}}{\phantom{e}}\!.\,. V_{n+\nu}|
+\cdots+
\delta_{m-n}|V_nV_{n+1}\dots \underset{\underset{k}{\uparrow}}{X_{m,n+k}}.\!
\overset{\overset{j}{\vee}}{\phantom{e}}\!.\,. V_{n+\nu}|\Bigr)\\
&
+
2w_m\Bigl(\delta_{m-n-1}|V_n\dots \underset{\underset{k+1}{\uparrow}}{Y_{m,n+k+1}}.\!
\overset{\overset{j}{\vee}}{\phantom{e}}\!.\,. V_{n+\nu}|
+\cdots+
\delta_{m-n+k-\nu}|V_n.\!
\overset{\overset{j}{\vee}}{\phantom{e}}\!.\,. Y_{m,n+\nu}|\Bigr).
\ea
\ee
Inspecting this last equation carefully we see that each term is quadratic in the fields 
$w$ and $\alpha^{(p)}$ for various values of $p$. Moreover, we see that there is no term 
of the form $\alpha^{(p)}\alpha^{(p)}$. However an almost
identical calculation applied to the Poisson bracket $\{\alpha^{(k)}_m,\alpha^{(k)}_n\}$
gives the following
\begin{proposition}
When represented in terms of the coordinates $\mathbf a$ on $\cW/G$, for each 
$k\in\{0,1,\dots,\nu-1\}$, the quadratic combination $\alpha^{(k)}\alpha^{(k)}$ appears in the 
entries of the Poisson tensor only in the bracket $\{\alpha^{(k)},\alpha^{(k)}\}$ and
\be\label{alkalkpb}
\ba
&\{\alpha^{(k)}_m,\alpha^{(k)}_n\}\\
&=
\left(2\sum_{l=1}^{\nu-k}\delta_{m-n-l} 
+ \sum_{l=0}^\nu{}_{l\neq k}\left[\sigma_{m-n-l}
+\sum_{r=0}^\nu{}_{r\neq k}\ \phi_{m-n+r-l}+\sum_{r=1}^\nu{}_{r\neq k}\ \delta_{m-n+r-l}
\right]\right)\alpha^{(k)}_m\alpha^{(k)}_n\\
&+\cdots
\ea
\ee
where the dots stand for quadratic terms of the form 
$(\alpha^{(p)}\cdot A_{m,n,p,q}\cdot\alpha^{(q)})$ for $p\neq q$, or 
$(\alpha^{(p)}\cdot B_{m,n,p}\cdot w)$, or $(w\cdot C_{m,n}\cdot w)$.
\end{proposition}

\begin{proof} The proof is by direct calculation. In fact the appropriate analogue of 
(\ref{kjpb}) differs in just one term: the last of the terms containing an $X$ is \\
$-2(-1)^{\nu-k}w_m\delta_{m-n+1}|V_nV_{n+1}\dots  \underset{\underset{k-1}{\uparrow}}{X_{m,n+k-1}}\dots\!\!\overset{\overset{j}{\vee}}{\phantom{e}}\!\!\dots V_{n+\nu}|$. 
The quadratic terms arise on the first line and amongst each of the terms containing a $Y$.
\end{proof}

We may now use the formulae (\ref{wwpb}, \ref{walkpb}, \ref{alkwpb}, \ref{alkalkpb}) and 
(\ref{defas})
to write down Poisson brackets involving the fields $a^{(k)}$, deciding in advance to discard all terms which are not quadratic in $a^{(k)}$.

Let us use the operator notation adapted to replacement of the functions $\sigma, \delta, \phi$ by suitable operators, for which the functions are the kernels. For a Poisson bracket relation of the form $\{x_m,y_n\}=x_m\cdot A_{m-n}\cdot y_n$ we denote
by $\widehat{\{x,y\}}$ the operator $A$ for which the function $m\mapsto A_m$ is the kernel. This notation may be used for 
the Poisson brackets $\{w,w\}$ and $\{w,\alpha^{(k)}\}$, but not for $\{\alpha^{(k)},\alpha^{(j)}\}$.
However we may define $\widehat{\{\alpha^{(k)},\alpha^{(k)}\}}$ to be the operator defining the 
quadratic term. The previous formulae are then written
\be\label{oppbs}
\ba
\widehat{\{w,w\}} &= \sum_{l=0}^{\nu-1}D^{-l}\sigma 
+ \sum_{l=0}^{\nu-1}\sum_{r=0}^{\nu-1}D^{r-l}\phi +
\sum_{l=0}^{\nu-1}\sum_{r=1}^{\nu-1}D^{r-l}\\
&=
\sum_{l=0}^{\nu-1}D^{-l}(\sigma-1)
+ \sum_{l=0}^{\nu-1}\sum_{r=0}^{\nu-1}D^{r-l}(\phi+1),\\
\widehat{\{w,\alpha^{(k)}\}}
&=
\sum_{l=0}^\nu{}_{l\neq k}D^{-l}(\sigma-1)
+ \sum_{l=0}^\nu{}_{l\neq k}\sum_{r=0}^{\nu-1}D^{r-l}(\phi+1),\\
\widehat{\{\alpha^{(k)},w\}}
&=
\sum_{l=0}^\nu{}_{l\neq k}D^{l}(\sigma+1)
+ \sum_{l=0}^\nu{}_{l\neq k}\sum_{r=0}^{\nu-1}D^{l-r}(\phi-1),\\
\widehat{\{\alpha^{(k)},\alpha^{(k)}\}}
&=
\sum_{l=0}^\nu{}_{l\neq k}D^{-l}(\sigma-1)
+ \sum_{l=0}^\nu{}_{l\neq k}\sum_{r=0}^\nu{}_{r\neq k}D^{r-l}(\phi+1)
+2\sum_{l=1}^{\nu-k}D^{-l}.
\ea
\ee
We have
\[
\{a^{(k)}_m, a^{(k)}_n\}
=w_m^{-1}w_n^{-1}\Bigl(\{\alpha^{(k)}_m,\alpha^{(k)}_n\}
- a^{(k)}_m\{w_m,\alpha^{(k)}_n\} - a^{(k)}_n\{\alpha^{(k)}_m,w_n\}
+ a^{(k)}_ma^{(k)}_n\{w_m,w_n\}\Bigr),
\]
which, after discarding non-quadratic terms, may be rewritten in the form
\[
\widehat{\{a^{(k)}, a^{(k)}\}}
=
\widehat{\{\alpha^{(k)},\alpha^{(k)}\}}
- \widehat{\{w,\alpha^{(k)}\}}
- \widehat{\{\alpha^{(k)},w\}}
+\widehat{\{w,w\}}.
\]
Now, using (\ref{oppbs})
we have
\[
\ba
\widehat{\{a^{(k)}, a^{(k)}\}}
&=
\sum_{l=0}^\nu{}_{l\neq k}D^{-l}(\sigma-1)
+ \sum_{l=0}^\nu{}_{l\neq k}\sum_{r=0}^\nu{}_{r\neq k}D^{r-l}(\phi+1) 
+ 2\sum_{l=1}^{\nu-k}D^{-l}\\
&+
\sum_{l=0}^{\nu-1}D^{-l}(\sigma-1)
+ \sum_{l=0}^{\nu-1}\sum_{r=0}^{\nu-1}D^{r-l}(\phi+1)\\
&-
\sum_{l=0}^\nu{}_{l\neq k}D^{-l}(\sigma-1)
- \sum_{l=0}^\nu{}_{l\neq k}\sum_{r=0}^{\nu-1}D^{r-l}(\phi+1)\\
&-
\sum_{l=0}^\nu{}_{l\neq k}D^{l}(\sigma+1)
- \sum_{l=0}^\nu{}_{l\neq k}\sum_{r=0}^{\nu-1}D^{l-r}(\phi-1).
\ea
\]
Distilling the above using the standard formula for geometric series, we obtain
\[
\widehat{\{a^{(k)}, a^{(k)}\}}
=
(D^{-\nu}-D^{-k})\Bigl[ 
(D^\nu - D^k)\phi + (D^\nu + D^k)
\Bigr].
\]
In other words
\[
\{a^{(k)},a^{(k)}\} = a^{(k)}
(D^{-\nu}-D^{-k})\Bigl[(D^{\nu}-D^{k})\phi+(D^{\nu}+D^k)\Bigr]
a^{(k)}
+ \cdots
\]
where the dots signify terms which are linear in the variables,
and the choice
\[
\phi=\frac{D^k+D^{\nu}}{D^k-D^{\nu}}
\]
forces the Poisson tensor to have linear dependence on $a^{(k)}$.

\noindent
On the other hand
\[
\{a^{(0)},a^{(k)}\}\ =\
\frac{1}{w}\Bigl(\{w',\alpha^{(k)}\}
-
\{w',w\}a^{(k)}-a^{(0)}\{w,\alpha^{(k)}\} + a^{(0)}\{w,w\}a^{(k)}\Bigr)\frac{1}{w}
\]
which may be rewritten in the form
\[
\ba
\widehat{\{a^{(0)},a^{(k)}\}}
&=\
(D-1)\widehat{\{w,\alpha^{(k)}\}} - (D-1)\widehat{\{w,w\}}\\
&=\
(D-1)\Bigl(\widehat{\{w,\alpha^{(k)}\}} - \widehat{\{w,w\}}\Bigr).
\ea
\]
Now, using (\ref{oppbs}), 
\[
\ba
\widehat{\{w,\alpha^{(k)}\}} - \widehat{\{w,w\}} &=\ \bigl(D^{-\nu} - D^{-k}\bigr)(\sigma-1)\
+\
\bigl(D^{-\nu} - D^{-k}\bigr)\sum_{r=0}^{\nu-1}D^r(\phi+1)\\
&=\
\bigl(D^{-\nu} - D^{-k}\bigr)(D-1)^{-1}\bigl[2 + (D^\nu-1)(\phi+1)\bigr]\\
&=\
\bigl(D^{-\nu} - D^{-k}\bigr)(D-1)^{-1}\bigl[ (D^\nu-1)\phi + (D^\nu+1)\bigr].
\ea
\]
Moreover
\[
\{a^{(0)}, a^{(0)}\} =\ \frac{1}{w}\Bigl(\{w',w'\}
-
\{w',w\}a^{(0)}-a^{(0)}\{w,w'\} + a^{(0)}\{w,w\}a^{(0)}\Bigr)\frac{1}{w}
\]
so that
\[
\ba
\widehat{\{a^{(0)}, a^{(0)}\}}
&=\
(2-D-D^{-1})\widehat{\{w,w\}}\\
&=\
-D^{-1}(D-1)^2\left[D^{1-\nu}\frac{D^\nu-1}{D-1}\frac{2}{D-1} 
+ D^{1-\nu}\left(\frac{D^\nu-1}{D-1}\right)^2(\phi+1)\right]\\
&=\
-(1-D^{-\nu})\bigl[2 + (D^\nu-1)(\phi+1)\bigr]\\
&=\
(D^{-\nu}-1)\bigl[(D^\nu-1)\phi+(D^\nu+1)\bigr].
\ea
\]
Hence the choice
\[
\phi=\frac{1+D^\nu}{1-D^\nu}
\]
forces $a^{(0)}$ to be a Casimir.


\end{document}